\documentclass[conference]{IEEEtran}

\usepackage{cite}      % Written by Donald Arseneau
\usepackage{graphicx}  % Written by David Carlisle and Sebastian Rahtz
\newcommand{\as}{$''$}
\newcommand{\seven}{$J=7\rightarrow6$}
\newcommand{\eight}{$J=8\rightarrow7$}
\newcommand{\nine}{$J=9\rightarrow8$}
\newcommand{\eleven}{$J=11\rightarrow10$}
\newcommand{\thirteen}{$J=13\rightarrow12$}
\newcommand{\coTW}{$^{12}$CO}
\newcommand{\coTH}{$^{13}$CO}
\newcommand{\czero}{[C~\small I\normalsize]}
\newcommand{\nplus}{[N~\small II\normalsize]}
\newcommand{\cplus}{[C~\small II\normalsize]}
\newcommand{\um}{$\mu$m}
\newcommand\apj{The Astrophysical Journal}%
\newcommand\apjl{The Astrophysical Journal}%
\newcommand\apjs{The Astrophysical Journal Supplement Series}%
\newcommand\aap{Astronomy \& Astrophysics}%
\newcommand\pasp{Pub. of the Astronomical Society of the Pacific}%
\newcommand\pasj{Pub. of the Astronomical Society of Japan}%
\begin{document}

\title{Observations in the 1.3 and 1.5~THz Atmospheric Windows with the Receiver Lab Telescope}

\author{\authorblockN{Daniel P. Marrone, Raymond
Blundell, \\ Edward Tong, Scott N. Paine, Denis Loudkov}
\authorblockA{Harvard-Smithsonian Center for Astrophysics \\
Email: dmarrone@cfa.harvard.edu}
\and
\authorblockN{Jonathan H. Kawamura}
\authorblockA{Jet Propulsion Laboratory, \\ 
California Institute of Technology}
\and
\authorblockN{Daniel L\"uhr, Claudio Barrientos}
\authorblockA{Universidad de Chile}}
\maketitle

\begin{abstract}
The Receiver Lab Telescope (RLT) is a ground-based terahertz
telescope; it is currently the only instrument producing astronomical
data between 1 and 2~THz. The capabilities of the RLT have been
expanding since observations began in late 2002. Initial observations
were limited to the 850~GHz and 1.03~THz windows due to the
availability of solid state local oscillators. In the last year we
have begun observations with new local oscillators for the 1.3 and
1.5~THz atmospheric windows. These oscillators provide access to the
{\eleven} and {\thirteen} lines of {\coTW} at 1.267 and 1.497~THz, as
well as the [N~II] line at 1.461~THz. We report on our first
measurements of these high CO transitions, which represent the
highest-frequency detections ever made from the ground. We also
present initial observations of [N~II] and discuss the implications of
this non-detection for the standard estimates of the strength of this
line.
\end{abstract}

\section{Introduction}
\label{s-intro}
\begin{figure}
\centering
\includegraphics[width=2.5in,angle=270]{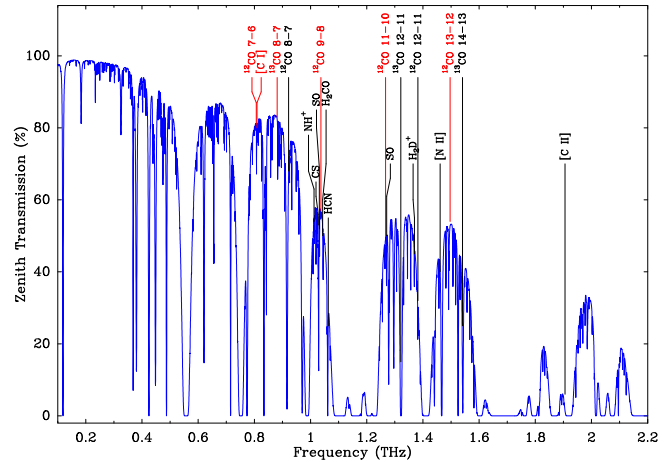}
\caption{Atmospheric transmission on Cerro Sairecabur on 24
January 2005 based on data from the Receiver Lab Fourier Transform
Spectrometer (FTS) \cite{PaineE00}. The FTS measures the sky emission
spectrum from 300~GHz to 3.5~THz at 3~GHz resolution; this spectrum is
then fit to an atmospheric model, which can be used to examine the
transmission at full resolution. The model indicates that at the time
of the measurement the precipitable water vapor (PWV) was only
93~\um. Several astronomically interesting lines are plotted for
reference, including those detected by the RLT (in red). The few
percent transmission at the 1.9~THz frequency of the [C~II] line is
unusual for this site, but suggests that even drier sites may provide
access to this important line from the ground.}
\label{f-tx}
\end{figure}

Atmospheric absorption prevents astronomical observations from the
ground at frequencies between 1 and 10~THz (300-30~\um), with the
dominant contributor to the opacity being tropospheric water
vapor. However, towards the ends of this frequency interval it is
possible to find atmospheric windows at very dry locations. In
particular, atmospheric transmission measurements between 1 and
3.5~THz show that a few strong windows open up under extremely dry
conditions \cite{MatsushitaE99,PaineE00}. An example of the
atmospheric transmission at a very dry site under the best conditions
is shown in Figure~\ref{f-tx}.

The Receiver Lab Telescope (RLT) is a ground-based terahertz
telescope, located 40~km north of the ALMA site in northern Chile. The
site, at an elevation of 5525 meters, shows some of the best terahertz
weather in the world, with transmission as high as 50\% observed in
three supra-terahertz windows in the last year. The RLT is equipped
with phonon-cooled HEB waveguide mixers for observations in four
atmospheric windows between 800~GHz and 1.6~THz. Within these windows
we have access to numerous atomic, molecular, and ionic lines,
including seven transitions of {\coTW} and {\coTH}, the 809~GHz
transition of {\czero}, and the 1.46~THz transition of {\nplus}. These
bright lines are relevant to many topics in astronomy including star
formation, the interstellar medium, and starburst/luminous infrared
galaxies. Other, weaker lines that are unique to terahertz astronomy
are also extremely interesting, in particular the 1.01~THz transitions
of NH$^+$, an undetected molecular ion in the formation chain of
ammonia, and the 1.37~THz ground-state transition of H$_2$D$^+$, a
tracer of the molecule responsible for chemistry inside cold molecular
cores. Astronomical interest in these and other lines has driven the
development of several instruments for ground-based terahertz
astronomy, as is discussed further in Section~\ref{s-future}. Due to
the atmospheric limitations at nearly all telescope sites, most of the
lines in the RLT bands have not been observed from the ground
(excepting, rarely, the 1.037~THz CO \nine\ line
\cite{PardoE01,KawamuraE02,MarroneE04apj}), and received little attention
from the Kuiper Airborne Observatory (KAO) before it was
decommissioned in 1995. Until the launch of Herschel in 2007-2008, or
possibly the installation of the first heterodyne instruments on the
Stratospheric Observatory For Infrared Astronomy (SOFIA, 2006-2007),
these lines will only be observable from ground-based telescopes like
the RLT or APEX.

The RLT and its first observations in the 1.03~THz window have been
described in previous editions of these proceedings and elsewhere
\cite{BlundellE02,MarroneE04,MarroneE04apj,MarroneE05}. Here we
discuss the first measurements made in the 1.3 and 1.5~THz windows,
the highest-frequency astronomical detections made from the ground at
radio frequencies, along with our first attempt at measuring the
\nplus\ line at 1.46~THz. 

\section{Observations at 1.3 and 1.5 THz}
\begin{figure}
\centering
\includegraphics[width=2.5in,angle=270]{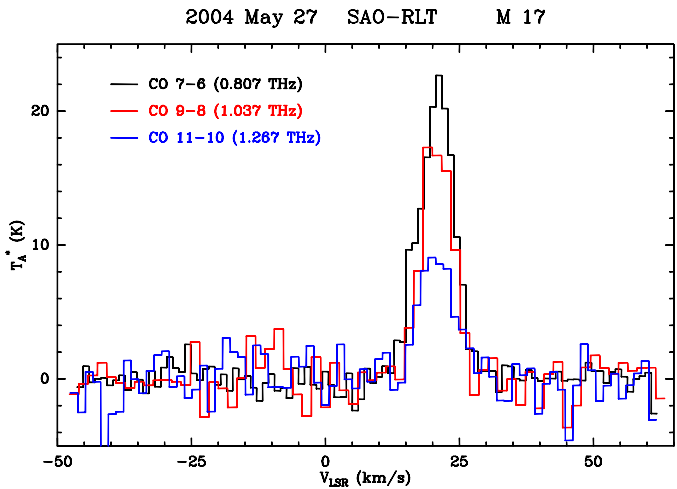}
\caption{CO emission detected in M17 from the RLT on 2004 May 27. At
the time of this measurement, the \coTW\ \eleven\ line was the highest
frequency line ever detected from the ground.}
\label{f-may27}
\end{figure}

In its first 18 months of operation the RLT was confined to
observations in the 850~GHz and 1.03~THz windows. For much of this
time we possessed a local oscillator (LO) source for the 1.3~THz
window, but were prevented from using it by the RF bandwidth of the
waveguide-coupled hot-electron bolometer mixer installed at the
telescope. In May 2004 we installed a mixer with slightly larger RF
bandwidth, sacrificing some performance in the low frequency windows
to enable operation at 1.3~THz. On May 27 we obtained the first
detection of an astronomical line in the 1.3~THz window, \coTW\
\eleven\ at 1.267~THz. This line, along with two lower transitions
observed in the same source on the same night, is shown in
Figure~\ref{f-may27}. All three lines have the same velocity extent,
as is expected for optically thick transitions, while the \eleven\
emission is weaker than the lower lines suggesting that the gas
temperature is not high enough to thermalize the 365~K $J=11$
rotational state.

Since this first observation, the RLT has routinely observed the CO
\eleven\ line in other sources. The atmospheric conditions on
Sairecabur allow regular observations of many high-frequency lines
(\coTW\ \seven, \nine, and \eleven, \coTH\ \eight, and \czero) that
are difficult or impossible to detect at other observatories. These
lines allow us to characterize the large-scale gas conditions in very
warm sources where the lower energy transitions available from other
telescopes are insensitive to the temperature.

\begin{figure}
\centering
\includegraphics[width=2.5in,angle=270]{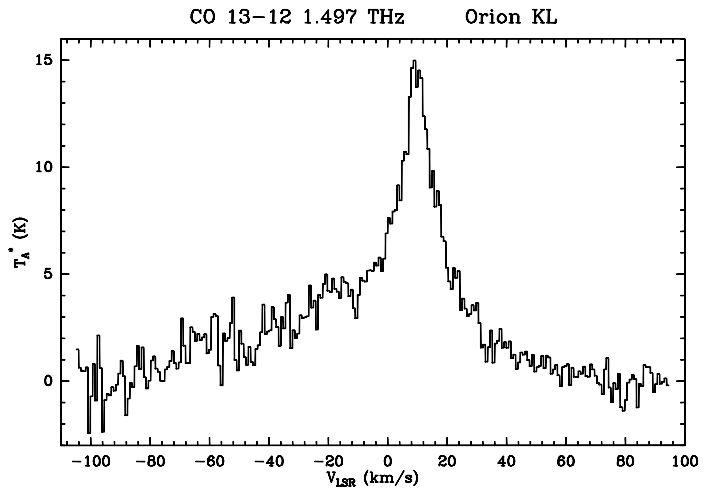}
\caption{\coTW\ \thirteen\ (1.497~THz) emission from Orion-KL, as
measured with the RLT on 2004 December 17. The flux scale is somewhat
uncertain because the telescope efficiency has not yet been measured
at this frequency, but the amplitude matches higher frequency
observations made with the KAO. This now stands as the highest
frequency line detected from the ground.}
\label{f-CO13}
\end{figure}

RLT observations moved to even higher frequency in December 2004 with
the arrival of an LO and receiver for 1.5~THz. The LO is on loan from
the Jet Propulsion Laboratory and was constructed as a prototype for
the HIFI instrument of the Herschel satellite
\cite{WardE03}. The receiver was built and tested in the Receiver Lab
and some of the testing is described elsewhere in these proceedings
\cite{TongE05}. As of this writing, only two marginal nights have been
available for observations at 1.5~THz with this receiver. Most of this
time was reserved for the \nplus\ line, but several minutes were spent
observing \coTW\ \thirteen\ at 1.497~THz to confirm that the receiver
was functioning properly. A detection of this line in Orion-KL, using
only 4 minutes on-source integration time, is shown in
Figure~\ref{f-CO13}. This detection represents the highest frequency
line measured from the ground and the only line observed in the
1.5~THz atmospheric window.

\section{Observations of \nplus}
The 1.4611~THz \nplus\ line is one of the most important targets of
ground-based terahertz astronomy. The FIRAS instrument
\cite{FixsenE94} on the COBE satellite, which mapped the entire sky at
low-angular and spectral resolution (7$^\circ$ beam, 5.4~GHz maximum
spectral resolution), found that the {\nplus} lines at 1.46 and
2.46~THz (205 and 122~\um) were the brightest lines in the Galaxy
after the 1.90~THz (157~\um) line of {\cplus}
\cite{WrightE91,BennettE94,FixsenE99}. These two lines can be used
together as a density probe for gas up to $\sim10^3$~cm$^{-3}$,
typical of the diffuse warm ionized medium \cite{KeenanE01}. The
higher-frequency {\nplus} line has been studied in this galaxy and
others at angular resolution comparable to that of the RLT (but much
lower velocity-resolution) using the Long-Wavelength Spectrometer on
the Infrared Space Observatory satellite \cite{ISO-LWS}. This
instrument was not sensitive to the 1.46~THz line and it is therefore
poorly studied: there are only two published detections, both from the
KAO \cite{ColganE93,PetuchowskiE94}.

Ground-based telescopes at exceptional locations like the South Pole
and the Atacama sites have access to this line in the 1.5~THz window,
although the transmission is somewhat degraded by a nearby strong
O$_2$ line at 1.4668~THz (at the line center, $\tau_{O_2}\simeq130$
for the South Pole and Sairecabur). The effect is worse at lower
altitude where pressure-broadening increases the O$_2$ line width; at
the South Pole it contributes an opacity of $\sim0.35$ at the {\nplus}
frequency, compared to $\sim0.16$ at Sairecabur\footnote{Based on
calculations performed with the {\it am} atmospheric model \cite{am},
available at http://cfarx6.cfa.harvard.edu/am}. With our new 1.5~THz
LO the RLT now has access to \nplus, and observations of this line are
now our key science goal.

\begin{figure}
\centering
\includegraphics[width=2.5in,angle=270]{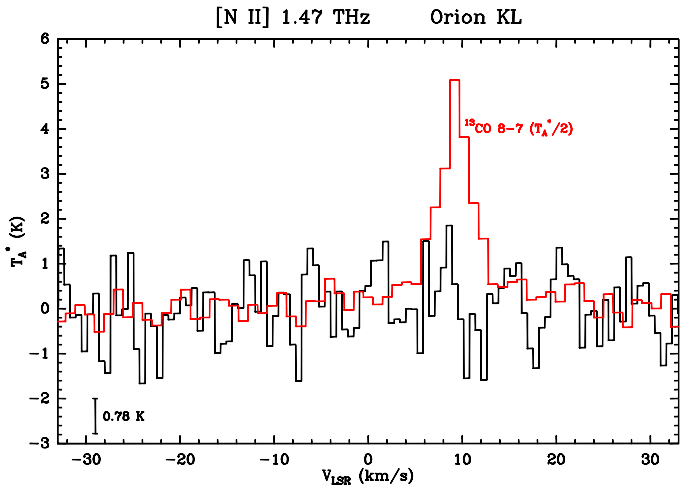}
\caption{[N~II] in Orion-KL, with {\coTH} {\eight} (0.881~THz)
overplotted as a rough velocity reference. The spectral rms is shown
in the lower left.}
\label{f-nii}
\end{figure}

As mentioned above, the 1.5~THz receiver arrived at the RLT in
December 2004 shortly before the end of the observing year. The two
nights available were somewhat below average for observations in this
window, with transmission of 11-13\% and 15-16\%, respectively, at
1.461~THz. Orion-KL was well placed in the sky for our observations
and contains an extended region of ionized gas, the edges of which
contain strongly excited CO (see Figure~\ref{f-CO13}), so we used this
as our main source. We also briefly attempted NGC~2024~IRS5 and
G270.3+0.8, using much less integration time and did not detect
\nplus\ emission. The resulting spectrum at the {\nplus} frequency is
shown in Figure~\ref{f-nii}, with {\coTH} {\eight} overplotted to
indicate the velocity extent of another optically thin line in this
source (although the CO emission traces slightly different gas). No
detection is apparent. The observations of Orion-KL totaled 78~minutes
on source and the rms on the spectrum is around 0.8~K, although the
telescope efficiency has not been measured at this frequency and could
be different from our (conservatively low) assumption. Higher
efficiencies would place even more stringent limits on the line
strength.

The lack of a detection of this line comes as something of a surprise
to us; many groups have proposed ambitious studies of {\nplus} and it
is expected to be quite bright. Of course, because there is little
data on its strength on angular scales smaller than the large COBE
beam one must make many assumptions to arrive at a predicted
strength. The simplest argument (and one that is frequently used) is
to take the {\cplus} line strength measured from the KAO and divide by
ten, the average of {\cplus/\nplus} as observed by COBE
\cite{WrightE91}. In the case of Orion-KL the velocity-resolved
{\cplus} observations of \cite{BoreikoE88} suggest a brightness
temperature of around 5~K, easily measured with our sensitivity. Our
non-detection suggests that this common argument is too simplistic. In
fact, the average over the whole sky is not representative of the
{\nplus} and {\cplus} emission in a given smaller region because the
two lines trace different gas. The {\nplus} line can be expected to be
present over much of the sky at a low level, while the {\cplus}
emission has a diffuse component but is most often found on the
surfaces of molecular clouds, which fill a much smaller fraction of
the sky. When averaged over the whole sky at low resolution, this
difference in filling factor suppresses the \cplus/\nplus\ ratio,
making this argument unreliable. A better estimate of the emission in
a small patch of sky can be made from KAO observations of a somewhat
analogous source, G333.6-0.2 \cite{ColganE93}. Both \cplus\ and
\nplus\ were detected in this source, with a line ratio of
\cplus/\nplus$_{1.46THz}=50$. Given this ratio, we may expect
something closer to 1~K in Orion-KL, which is entirely consistent with
our observations.

RLT observations have been suspended since January for the summer wet
season known locally as ``Bolivian Winter''. Operations resume in late
April or early May with a new list of target sources. In particular,
we are using ISO observations of the 2.46~THz line of {\nplus} to
select our sources. Although COBE observations suggest that the
Galactic average {\nplus}$_{1.46}$/{\nplus}$_{2.46}$ line ratio is
approximately unity \cite{FixsenE99}, in the individual sources we
observe it is likely to be lower. In the high-density limit
($n>10^3$~cm$^{-3}$) this ratio is around 0.1, and most gas in
discrete sources will be at or above this density threshold. The ISO
observations are not velocity resolved in most sources so the measured
line fluxes cannot be directly inverted to obtain a peak line
strength, but many sources with {\nplus} emission stronger than that
observed in Orion have been obtained in the appropriate hour angle
range.

\section{Prospects for THz Astronomy from the Ground}
\label{s-future}
For the next two or more years, terahertz astronomy will only be
possible from the ground. The Receiver Lab Telescope has now
demonstrated observations of astronomical line radiation in all three
of the atmospheric windows between 1.0 and 1.6~THz. From our site and
nearby sites in northern Chile, the 1.5~THz window is likely to be the
highest frequency window that will be regularly usable for
astronomy. Observations of the transmission on very dry nights
(multiple instances of PWV below 200~\um, including the 93~\um\ shown
in Figure~\ref{f-tx}, have been been observed in the last year)
suggest that from an even drier site one may be able to move to higher
frequencies, including observations of the \cplus\ line at
1.9~THz. Proposed observations from Antarctic Dome A, for which
measurements of submillimeter opacity are not yet available, may be
able to make this step to higher frequencies if PWV predictions for
this site are accurate.

Ground-based measurements will continue to have an important place in
terahertz astronomy even in the era of SOFIA and Herschel. First,
telescopes on the ground can be much larger than is possible from
airplanes or from space at similar cost. In the next year the 12-meter
APEX telescope will begin observations above 1~THz with multiple
receivers. Planning for an even larger telescope, the 25-meter
Caltech-Cornell Atacama Telescope (CCAT), are underway. CCAT would
achieve angular resolution around 2\as\ at 1.5~THz, better than is now
available from any single-aperture radio telescope, and almost an
order of magnitude better than will be obtained from SOFIA or
Herschel. Moreover, Herschel will lack receiver coverage between 1.25
and 1.41~THz, a gap that lines up well with the 1.3~THz atmospheric
window. Sensitive observations from the ground, particularly with the
angular resolution available from these larger telescopes, will access
important science that Herschel will miss.

\section*{Acknowledgment}
The authors thank the many current and former members of the Receiver
Lab who have worked on bringing the RLT into existence and contributed
to its operations in Chile, in particular, Hugh Gibson and Cosmo
Papa. Site evaluation and RLT development could not have taken place
without the ever present support of Irwin Shapiro, former director of
the Center for Astrophysics. The project also owes a great deal to the
many years of assistance provided by Jorge May and Leo Bronfman at
Universidad de Chile. We also thank Imran Mehdi, John Ward, and their
LO team for providing us with an oscillator for 1.5~THz
observations. This work was performed for the Jet Propulsion
Laboratory, California Institute of Technology, sponsored by the
National Aeronautics and Space Administration. DPM acknowledges
support from an NSF Graduate Research Fellowship.

\end{document}